\documentclass[preprint,aps,prl,showpacs]{revtex4-1}

\usepackage{graphicx}  
\usepackage{dcolumn}   
\usepackage{amsmath}

\usepackage[utf8]{inputenc}   
\usepackage[T1]{fontenc}      
\usepackage{textcomp}

\begin{document}

\title{Crystalline Soda Can Metamaterial exhibiting Graphene-like Dispersion at subwavelength scale}
\author{S. Yves, F. Lemoult, M. Fink, G. Lerosey}
\email[]{geoffroy.lerosey@espci.fr}
\affiliation{Institut Langevin, CNRS UMR 7587, ESPCI Paris, PSL Research University, 1 rue Jussieu, 75005 Paris, France}
\date{\today}

\begin{abstract}
Graphene, a honeycomb lattice of carbon atoms ruled by tight-binding interaction, exhibits extraordinary electronic properties due to the presence of Dirac cones within its band structure. These intriguing singularities have naturally motivated the discovery of their classical analogues. In this work, we present a general and direct procedure to reproduce the peculiar physics of graphene within a very simple acoustic metamaterial: a double lattice of soda cans resonant at two different frequencies. The first triangular sub-lattice generates a bandgap at low frequency, which induces a tight-binding coupling between the resonant defects of the second Honeycomb one, hence allowing us to obtain a graphene-like band structure. We prove the relevance of this approach by showing that both  numerical and experimental dispersion relations exhibit the requested Dirac cone. We also demonstrate the straightforward monitoring of the coupling strength within the crystal of resonant defects. This work shows that crystalline metamaterials are  very promising candidates to investigate tantalizing solid-state physics phenomena with classical waves.
\end{abstract}

\maketitle

For the last decades a lot of attention has been paid to bi-dimensional materials, such as graphene, for their extraordinary electronic properties. The presence of a Dirac cone in the band structure gives them the electronic properties of zero gap semi-conductors which are promised to many applications. Besides, it is also responsible for new physical phenomena such as the bosonic propagation of fermions or the Zitterbewegung effect~\cite{Novoselov2005,Katsnelson2006,Geim2007,RevModPhys.81.109,RevModPhys.80.1337,RevModPhys.82.2673}. Even if the fabrication of such materials have really progressed and it becomes now possible to have very pure samples, it remains difficult to probe locally the electronic properties at the atomic scale. Therefore, it becomes interesting to emulate those physics at a higher scale with photons or phonons, thanks to the analogy between electronic band structures and the propagation bands developed in the context of photonic (phononic) crystals. In this way, there have been several demonstrations of the existence of a Dirac cone in two dimensional photonic (phononic) crystals~\cite{PhysRevLett.101.264303,Bittner2010,PhysRevB.85.064301,PhysRevB.86.035141}. However, these examples present a significant difference with the graphene: the triangular lattice with solely one scatterer per unit cell allows the existence of the Dirac cone thanks to multiple scattering of waves. Other examples exhibiting two resonant scatterers within the unit cell demonstrate also the occurrence of Dirac dispersion~\cite{PhysRevB.80.155103,zhong2011acoustic,PhysRevLett.108.174301,PhysRevB.87.115143}. Nevertheless, in these cases the degeneracy is solely due to the geometry of the honeycomb lattice, but they lack the tight-binding interaction required by the graphene model. This severely restricts the analogy between the two systems and make any conclusions hard to transpose from the photonic (phononic) examples to the electronic properties of graphene.

Graphene's electronic band structure is due to the following two features: its honeycomb structure and the nearest-neighbor interaction within carbon atoms described by a tight-binding model~\cite{PhysRev.71.622,PhysRevB.66.035412}. Therefore, a macroscopic scale photonic (phononic) analogue should share these features which is for example the case for Mie resonators placed on a ground plane~\cite{PhysRevB.82.094308,PhysRevB.88.115437}, coupled cavities in photonic crystals~\cite{PhysRevB.84.075477}, or plasmonic nanostructures~\cite{PhysRevLett.102.123904}. In this letter, we present a general scheme to mimic tight-binding governed Hamiltonians, at the deep subwavelength scale using locally resonant metamaterials. To do so, we first design a trangular sub-lattice of resonators that creates a full bandgap. The latter allows us to introduce a second sub-lattice of resonant defects organized on the required honeycomb structure. The obtained defects hence behave as subwavelength cavities that can solely interact through evanescent coupling. This, in turns, realizes a deep subwavelength honeycomb crystal of defects with a nearest-neighbor interaction very similar to the Hamiltonian of graphene. This is experimentally demonstrated with a very simple acoustic metamaterial: a lattice of soda cans. Moreover, since the frequency detuning of the defects is simple to realize experimentally, our system directly permits to investigate the effect of the strength of interaction between neighboring defects without changing the lattice dimensions. This demonstrates that engineering locally resonant metamaterials at a subwavelength scale makes them a straightforward tabletop platform for the study of exciting solid-state physics.

\begin{figure}[bt]
	\begin{center}
\includegraphics [scale=1] {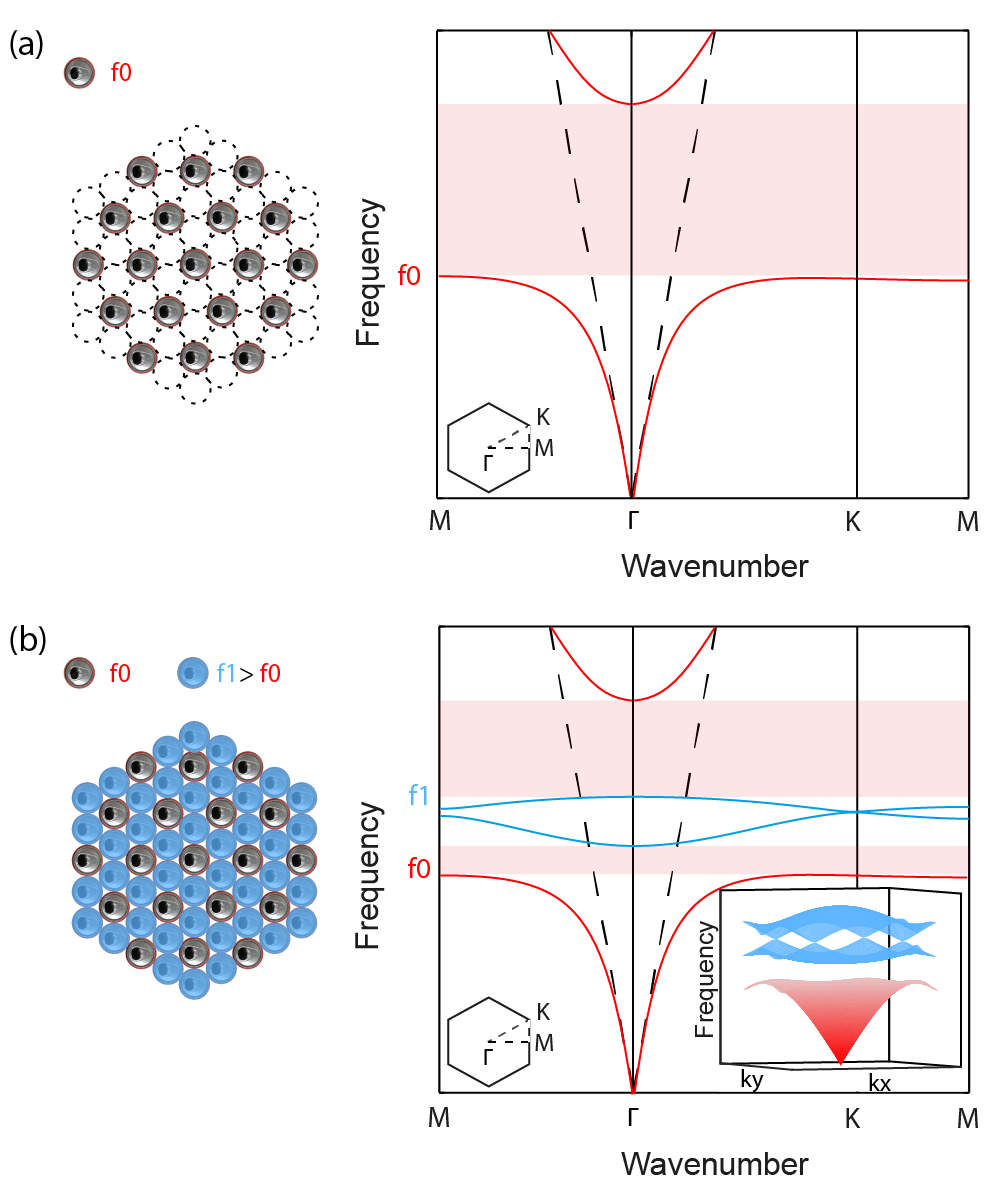}
	\caption{\label{Fig1} Principle of the study. (a) Dispersion relation of a triangular lattice of empty cans resonating at the frequency $f_0$. It is a polariton (red) that creates a bandgap (red shaded area). (b) Addition of a honeycomb sublattice of partly filled cans of higher resonance frequency $f_1$ creates two bands (blue) inside the bandgap. They cross at the K point of the Brillouin zone forming a single point degeneracy known as a Dirac point (inset).}
	\end{center}
\end{figure}

Acoustic metamaterials are man-made composite media which are by definition structured at scales that are much smaller than the sound's wavelength of operation in the surrounding medium. Therefore, these systems are usually described by macroscopic effective properties, which can be tailored at a mesoscopic scale in order to manipulate the sound propagation~\cite{Engheta2006,deymier2013acoustic,Cummer2016,Mae1501595}. Hence, several intriguing features have been demonstrated with these systems: negative refraction and acoustic cloaking~\cite{PhysRevE.70.055602,PhysRevLett.110.175501,Zigoneanu2014,PhysRevLett.112.144301,PhysRevLett.108.124301}, or density near-zero metamaterials~\cite{fleury2013extraordinary}, 
or deep-subwavelength imaging~\cite{Zhu2011} for example. A specific class of these acoustic systems are the locally resonant metamaterials whose unit cell presents a resonant scattering cross-section~\cite{Liu1734,PhysRevE.91.023204,Romero-Garcia2016}. In this study, we use soda cans as low-loss subwavelength acoustic Helmholtz resonators~\cite{PhysRevLett.107.064301,Lemoult2013,Kaina2015,Lemoult_2016}. Their resonance frequency depends on the volume of air inside the can and is around 420~Hz ($\lambda$=80~cm) if empty. This resonant wavelength is far larger than the size of the can which makes it a relevant building block for our acoustic metamaterial. As a wave propagates in this particular medium, interference phenomena lead to an hybridization with the resonant elements making the wave behaves as a polariton. Hence, below the resonance frequency $f_0$ of a single empty can, the system presents propagating modes that are more and more subwavelength approaching the resonance from below~\cite{PhysRevLett.107.064301}. On the contrary, above $f_0$ wave propagation in the lattice is forbidden within a certain frequency range. This is a hybridization bandgap, resulting from the resonant nature of the cans and independent of their spatial order~\cite{sigalas1993band}. We can then introduce defects by adding resonators whose resonance frequency $f_1$, higher than $f_0$, falls within the bandgap frequency range. Such resonant defects can only couple one to another through evanescent decay since the bandgap prevents the existence of propagating modes, as it has been demonstrated in~\cite{Lemoult2013} while building subwavelength waveguides. 

\begin{figure}[bt]
	\begin{center}
\includegraphics [scale=1] {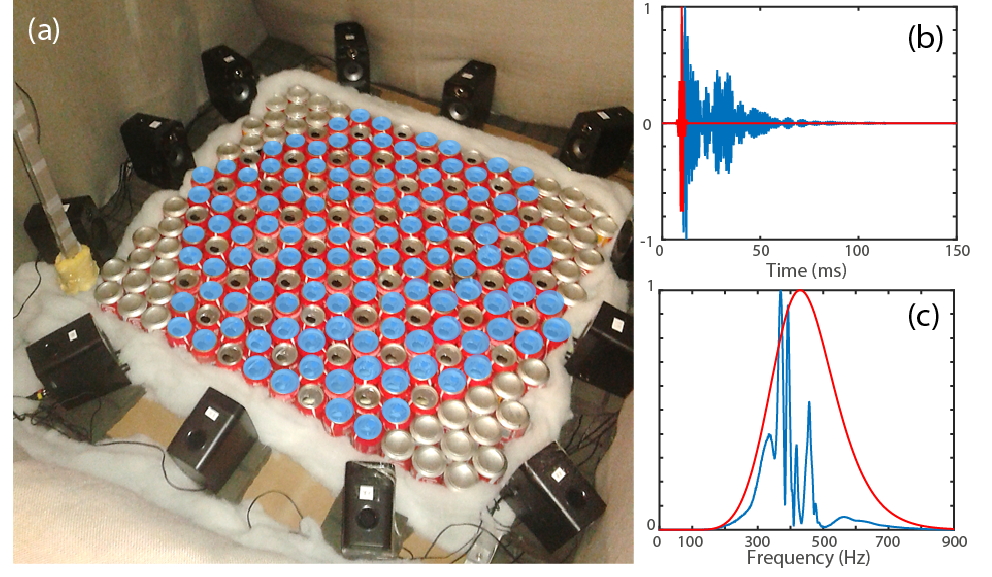}
	\caption{\label{FigExp} (a) Experimental setup. Shaded blue cans are partly filled with water. (b) We sent a chirp (red) with the speakers and record the signal with a microphone above the sample (blue). (c) Spectra of (b).}
	\end{center}
\end{figure}

In order to leave room for organizing the defects along a honeycomb arrangement later on, we first build a non compact triangular lattice of empty cans (Fig.~\ref{Fig1}.a).
By simulating the unit cell with Bloch boundary conditions, we extract the dispersion relation in the principal directions of the crystal. It exhibits the polariton behavior previously described: a full bandgap exists above the resonance frequency $f_0$ of a single empty can. A honeycomb lattice of defects, made of cans whose resonant volume is reduced by pouring some water in it, is then placed in the interstices of the previous triangular sublattice (Fig.~\ref{Fig1}.b). If the medium is excited around the resonance frequency $f_1$ of the partly filled cans, the bandgap prevents the acoustic waves to propagate and the acoustic field is confined on the filled cans. However, the field is able to tunnel from one defect to another which induces the desired tight-binding coupling, fulfilling the conditions to rigorously reproduce the physics described by the graphene Hamiltonian at a macroscopic scale. The resulting dispersion relations obtained for this new medium (Fig.~\ref{Fig1}.b) exhibit two new propagating bands, represented in blue, laying in the bandgap of the previous crystal around the frequency $f_1$. Those blue bands present a single point degeneracy at each K point of the first Brillouin zone (see inset), known as a Dirac cone, which is a feature of the honeycomb lattice and graphene. 

\begin{figure*}[hbt]
	\begin{center}
\includegraphics[scale=1]{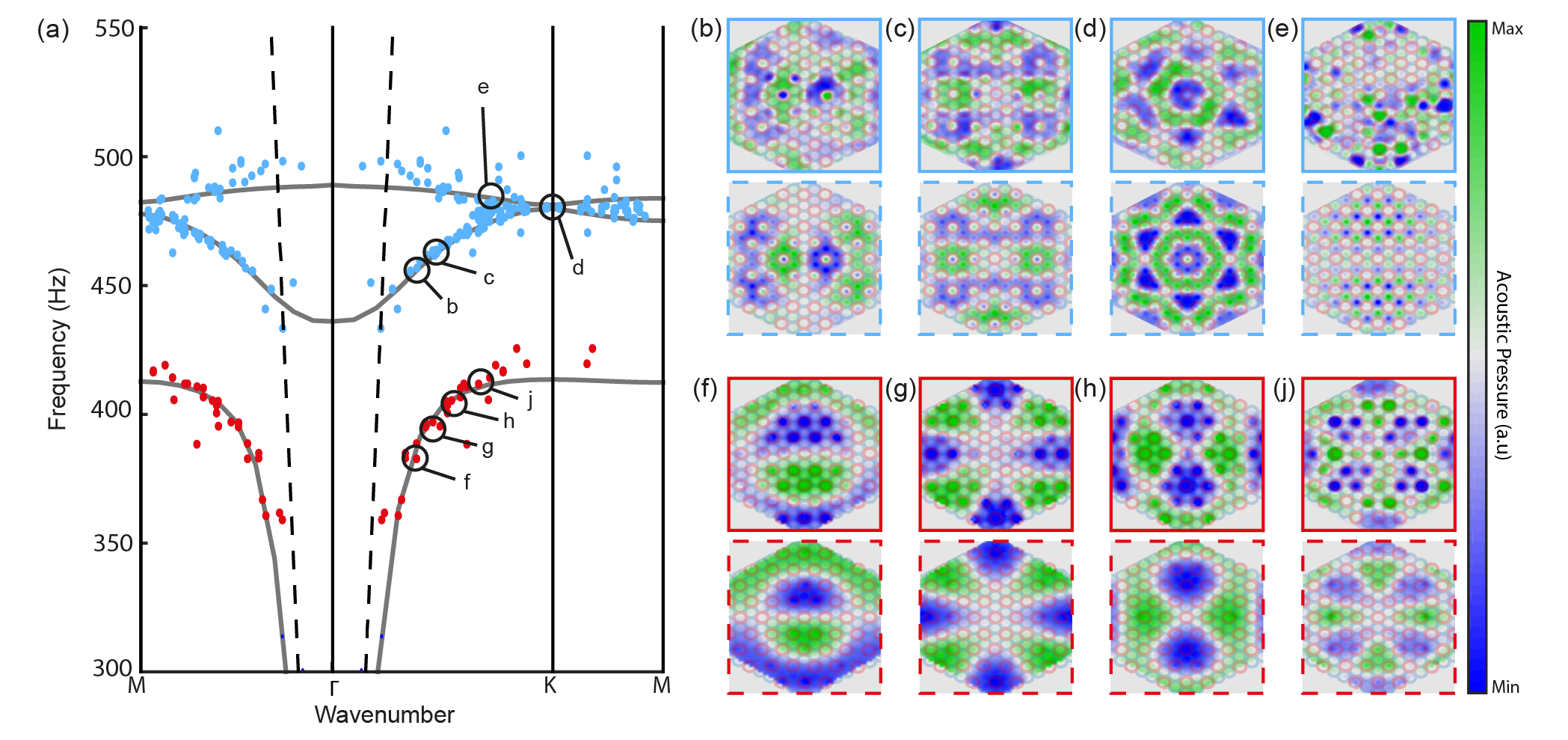}
	\caption{\label{FigDiracCone} (a)Experimental dispersion relation superimposed on numerical results (gray curve) corresponding to an experiment made with 8cL of water in the defects. We can see the polaritonic modes (red points) and the two defect bands (blue points) that cross at the K point. (b-e) Corresponding calculated (dashed line boxes) and experimental modes (continuous line boxes) of the defect bands. The cans (blue for the filled ones and red for the empty ones) are superimposed on the maps. (f-j) Same as (b-e) but for the polaritonic modes. }
	\end{center}
\end{figure*}

After this numerical proof of concept, we implement the experimental version of the metamaterial composed of simple soda cans by realizing a medium made of 169 cans, 47 being empty and 122 filled with 8~cL of waters (Fig.~\ref{FigExp}.a). The finite sized macroscopic medium has a shape of an hexagon in order to respect the symmetry of the unit cell and all edges are identical. We place the medium in an anechoic chamber and excite it with an array of twelve commercial loudspeakers positioned all around. They are located approximately 15~cm far from the sample and directed along the 6 symmetrical axes of the large hexagon in order to scan all the directions. A microphone, mounted on a two-dimensional translational stage, measures the acoustic field above the metamaterial. Using multiple sources and recording the received signal for each emission gives us the opportunity to create complex excitation patterns thanks to the linearity of the wave equation. This is carried out as a post-processing step by combining different maps. A single emission consists of a 10~s chirp that spans the frequency range from 100~Hz to 900~Hz. By applying match-filtering, it amounts to emit a 1.4~ms gaussian pulse (Fig.~\ref{FigExp}.b) centered on 450~Hz. A typical match-filtered signal measured at an arbitrary location in the middle of the scanning area above the hexagon is shown in blue: the signal spans for longer time than the initial pulse as a signature of stationary modes created in the finite-sized hexagon. Its frequency contents (Fig.~\ref{FigExp}.c) exhibits the presence of a first set of peaks below 450~Hz followed by a bandgap. Then other peaks appears at frequencies below 500~Hz, which corresponds to the defects' bands. Nevertheless, at this stage, it is not enough to conclude on the fact that we achieved to obtain the band structure of graphene within our system.

In order to effectively prove that this metamaterial behaves like an acoustic analog of graphene, we will extract the dispersion relation from the experiment thanks to the two dimensional mapping. A temporal Fourier transform of our signals gives us field maps for each frequency contained in the emitted signal. We need now to relate these spatial modes to a wavenumber. However using a direct 2D spatial Fourier transform of those field maps is not sufficient. Indeed, even though soda cans present acceptable amounts of dissipation, their intrinsic viscous and thermic losses make the most subwavelength modes to spectrally overlap and prevent us from attributing a precise wavenumber to a given frequency. This is even more problematic, given that we try to observe the Dirac cone, key feature of the graphene band structure, which is at the border of the first Brillouin zone. To overcome this difficulty we implement a more complex data treatment which consists in comparing measured field patterns to those calculated thanks to Comsol simulations without losses.
  
This procedure leads to the results summarized in Fig.~\ref{FigDiracCone}. For each frequency the spatial map corresponding to the right excitation pattern is compared to all of the numerical maps, and a wavenumber is attributed according to the numerical dispersion relation. For example, at 380~Hz the measured spatial pattern of Fig.~\ref{FigDiracCone}.f matches well with the numerical spatial pattern represented below which permits to attribute a point in the principal directions of the first Brillouin zone. Accumulating such maps' correspondences permit to draw the cloud of points shown in the Fig.\ref{FigDiracCone}.a which well fits the simulated dispersion relation (gray curve). Noticeably, this clearly exhibits the existence of the two bands crossing at K point. In the other panels, we show different matching examples of the experimental maps (continuous line box) above their corresponding numerical counterparts (dashed line box). In spite of minor differences due to losses inherent to the experiment, notably in the case of the modes of the folded blue band, we notice a good agreement between them. 

A closer look at these spatial modes is needed to fully understand the dispersion relation. Below 420~Hz, the field is mostly distributed on the triangular lattice of empty cans. The attributed red points exhibit the expected polaritonic behavior with a flat asymptote near $f_0=420$~Hz. Above 440~Hz, the experimental pressure is concentrated on the defect cans arranged with the honeycomb scheme. The blue points associated to those defect modes fall within the polaritonic bandgap and clearly exhibit the two desired bands crossing at K. 

At this stage, we can conclude that we have managed to create a honeycomb crystal of coupled defects in the bandgap created by a triangular lattice of empty cans. Nevertheless, this experimental configuration does not show the exact equivalent of the electronic band structure of a tight-binding model of graphene. Indeed, although presenting a Dirac cone at the K point of the Brillouin zone, the defect bands do not have a symmetrical profile about this single point degeneracy: the lower band seems broader than the upper one.

To obtain this localized interaction, we take advantage of another property of the hybridization bandgap: the frequency dependence of its efficiency in attenuating wave propagation~\cite{Lemoult2013}. For instance, at frequencies just above $f_0$, the bandgap is very efficient. In this way, waves are strongly confined on defects if the resonance frequencies $f_1$ is only slightly detuned from $f_0$. This enhanced confinement is responsible for harder tunneling between neighboring defects, finally resulting in a more rigorous tight-binding interaction. In practice, this corresponds to a configuration in which cans are filled with a smaller amount of water. 

\begin{figure*}[hbt]
	\begin{center}
\includegraphics[width=15cm] {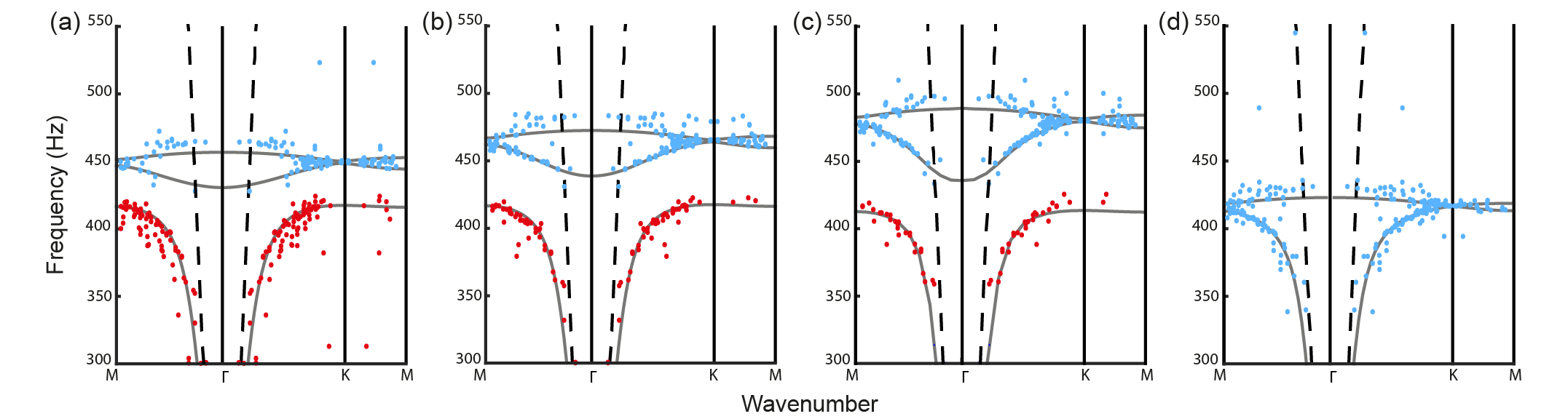}
	\caption{(a) Experimental dispersion relations for defects made of cans with 4 cL of water inside. (b) Same as (a) but for 6 cL. (c) Same as (a) but for 8 cL (c). (d). Experimental dispersion relation of a honeycomb lattice of empty cans, outside any bandgap.}
	\label{Fig4}
	\end{center}
\end{figure*}

On that account, the same experiment as before is carried but with less water inside the cans (4~cL). The obtained dispersion relation (Fig.~4.a), although still affected by the losses inherent to the experiment, presents the desired symmetrical profile around the Dirac cone degeneracy at K point. This experimental configuration is therefore accurately described by the tight-binding Hamiltonian as it is performed for the electronic band structure of graphene. We have therefore managed to obtain with our acoustic resonant metamaterial a rigorous macroscopic acoustic equivalent of graphene.

We then go further and carry out other experiments increasing the level of water in the cans, {\it ie.} the detuning of $f_1$ above the resonance frequency $f_0$ creating the bandgap. This means that the confinement distance created by the triangular lattice of empty cans gradually decreases. In this way, the coupling between the defects tends to be less and less of nearest-neighbor type but the coupling is extended to more remote defects. This is traduced in the shape of the experimental dispersion relation obtained for 6~cL (Fig.~\ref{Fig4}.b) which clearly shows an asymmetry: the lower band spreads more towards the lower frequencies. Increasing again the amount of water inside the cans, we continuously modify the nature of the interaction within the honeycomb lattice and the detuning becomes too high so that the filled cans barely resonate within the bandgap created by the empty ones: a transition from a tight-binding type of coupling to a polaritonic one occurs. Interestingly, for 8~cL of waters the dispersion relation still exhibits a Dirac degeneracy (Fig.~\ref{Fig4}.c) but at a frequency near the upper edge of the bandgap calculated in Fig.~\ref{Fig1}.a. This means that even with a polaritonic type of coupling the presence of two resonant ``atoms'' per unit cell is sufficient to create it, even if it is not a complete equivalent of the graphene physics. 

To be convinced of this latter assertion, we also carried an experiment with a resonant metamaterial made of solely empty cans but organized as an honeycomb array. The obtained dispersion relation (Fig.~\ref{Fig4}.d) also exhibits the Dirac degeneracy at K point but this occurs for a frequency within the initial polaritonic band of the non-compact triangular lattice of empty cans. This time, the existence of the extra-band is solely due to the presence of the two atoms per unit cell, in a similar way as~\cite{PhysRevB.80.155103,zhong2011acoustic,PhysRevLett.108.174301,PhysRevB.87.115143}. It can therefore be interpreted as a branch created by tight-binded dipolar resonators (each one made by two cans) tunneling through the monopolar resonance of two in-phase resonating cans. This is a clear signature that an homogenization procedure for describing the propagation in a metamaterial is not sufficient, as we have already pointed out in~\cite{Kaina2015}. It also explains why the previously observed transition from tight-binding to polaritonic coupling does not affect the existence of the Dirac degeneracy and the shape of the upper band but only the symmetry with respect to the Dirac frequency. 
 
To conclude, in this Letter we have shown that monitoring the structure and the composition of an acoustic resonant metamaterial enables to create physics that cannot be seen from a homogenization point of view, although the dimensions are far below the wavelength. In this manner, we have demonstrated that it is possible to design a macroscopic acoustic analog of graphene using simple soda cans. This analogy becomes true solely in some conditions that we have described, namely when the bandgap is sufficiently attenuating to limit the interaction to the nearest neighbors. The crystalline approach of these resonant metamaterials we have adopted permits to understand the local interactions and makes this class of media a promising platform to investigate solid state physics phenomena at a macroscopic scale.

\bibliography{biblio}

\begin{thebibliography}{39}%
\makeatletter
\providecommand \@ifxundefined [1]{%
 \@ifx{#1\undefined}
}%
\providecommand \@ifnum [1]{%
 \ifnum #1\expandafter \@firstoftwo
 \else \expandafter \@secondoftwo
 \fi
}%
\providecommand \@ifx [1]{%
 \ifx #1\expandafter \@firstoftwo
 \else \expandafter \@secondoftwo
 \fi
}%
\providecommand \natexlab [1]{#1}%
\providecommand \enquote  [1]{``#1''}%
\providecommand \bibnamefont  [1]{#1}%
\providecommand \bibfnamefont [1]{#1}%
\providecommand \citenamefont [1]{#1}%
\providecommand \href@noop [0]{\@secondoftwo}%
\providecommand \href [0]{\begingroup \@sanitize@url \@href}%
\providecommand \@href[1]{\@@startlink{#1}\@@href}%
\providecommand \@@href[1]{\endgroup#1\@@endlink}%
\providecommand \@sanitize@url [0]{\catcode `\\12\catcode `\$12\catcode
  `\&12\catcode `\#12\catcode `\^12\catcode `\_12\catcode `\%12\relax}%
\providecommand \@@startlink[1]{}%
\providecommand \@@endlink[0]{}%
\providecommand \url  [0]{\begingroup\@sanitize@url \@url }%
\providecommand \@url [1]{\endgroup\@href {#1}{\urlprefix }}%
\providecommand \urlprefix  [0]{URL }%
\providecommand \Eprint [0]{\href }%
\providecommand \doibase [0]{http://dx.doi.org/}%
\providecommand \selectlanguage [0]{\@gobble}%
\providecommand \bibinfo  [0]{\@secondoftwo}%
\providecommand \bibfield  [0]{\@secondoftwo}%
\providecommand \translation [1]{[#1]}%
\providecommand \BibitemOpen [0]{}%
\providecommand \bibitemStop [0]{}%
\providecommand \bibitemNoStop [0]{.\EOS\space}%
\providecommand \EOS [0]{\spacefactor3000\relax}%
\providecommand \BibitemShut  [1]{\csname bibitem#1\endcsname}%
\let\auto@bib@innerbib\@empty
\bibitem [{\citenamefont {Novoselov}\ \emph {et~al.}(2005)\citenamefont
  {Novoselov}, \citenamefont {Geim}, \citenamefont {Morozov}, \citenamefont
  {Jiang}, \citenamefont {Katsnelson}, \citenamefont {Grigorieva},
  \citenamefont {Dubonos},\ and\ \citenamefont {Firsov}}]{Novoselov2005}%
  \BibitemOpen
  \bibfield  {author} {\bibinfo {author} {\bibfnamefont {K.~S.}\ \bibnamefont
  {Novoselov}}, \bibinfo {author} {\bibfnamefont {A.~K.}\ \bibnamefont {Geim}},
  \bibinfo {author} {\bibfnamefont {S.~V.}\ \bibnamefont {Morozov}}, \bibinfo
  {author} {\bibfnamefont {D.}~\bibnamefont {Jiang}}, \bibinfo {author}
  {\bibfnamefont {M.~I.}\ \bibnamefont {Katsnelson}}, \bibinfo {author}
  {\bibfnamefont {I.~V.}\ \bibnamefont {Grigorieva}}, \bibinfo {author}
  {\bibfnamefont {S.~V.}\ \bibnamefont {Dubonos}}, \ and\ \bibinfo {author}
  {\bibfnamefont {A.~A.}\ \bibnamefont {Firsov}},\ }\href {\doibase
  10.1038/nature04233} {\bibfield  {journal} {\bibinfo  {journal} {Nature}\
  }\textbf {\bibinfo {volume} {438}},\ \bibinfo {pages} {197} (\bibinfo {year}
  {2005})}\BibitemShut {NoStop}%
\bibitem [{\citenamefont {Katsnelson}\ \emph {et~al.}(2006)\citenamefont
  {Katsnelson}, \citenamefont {Novoselov},\ and\ \citenamefont
  {Geim}}]{Katsnelson2006}%
  \BibitemOpen
  \bibfield  {author} {\bibinfo {author} {\bibfnamefont {M.~I.}\ \bibnamefont
  {Katsnelson}}, \bibinfo {author} {\bibfnamefont {K.~S.}\ \bibnamefont
  {Novoselov}}, \ and\ \bibinfo {author} {\bibfnamefont {A.~K.}\ \bibnamefont
  {Geim}},\ }\href {\doibase 10.1038/nphys384} {\bibfield  {journal} {\bibinfo
  {journal} {Nat Phys}\ }\textbf {\bibinfo {volume} {2}},\ \bibinfo {pages}
  {620} (\bibinfo {year} {2006})}\BibitemShut {NoStop}%
\bibitem [{\citenamefont {Geim}\ and\ \citenamefont
  {Novoselov}(2007)}]{Geim2007}%
  \BibitemOpen
  \bibfield  {author} {\bibinfo {author} {\bibfnamefont {A.~K.}\ \bibnamefont
  {Geim}}\ and\ \bibinfo {author} {\bibfnamefont {K.~S.}\ \bibnamefont
  {Novoselov}},\ }\href@noop {} {\bibfield  {journal} {\bibinfo  {journal}
  {Nat. Mater.}\ }\textbf {\bibinfo {volume} {6}},\ \bibinfo {pages} {183}
  (\bibinfo {year} {2007})}\BibitemShut {NoStop}%
\bibitem [{\citenamefont {Castro~Neto}\ \emph {et~al.}(2009)\citenamefont
  {Castro~Neto}, \citenamefont {Guinea}, \citenamefont {Peres}, \citenamefont
  {Novoselov},\ and\ \citenamefont {Geim}}]{RevModPhys.81.109}%
  \BibitemOpen
  \bibfield  {author} {\bibinfo {author} {\bibfnamefont {A.~H.}\ \bibnamefont
  {Castro~Neto}}, \bibinfo {author} {\bibfnamefont {F.}~\bibnamefont {Guinea}},
  \bibinfo {author} {\bibfnamefont {N.~M.~R.}\ \bibnamefont {Peres}}, \bibinfo
  {author} {\bibfnamefont {K.~S.}\ \bibnamefont {Novoselov}}, \ and\ \bibinfo
  {author} {\bibfnamefont {A.~K.}\ \bibnamefont {Geim}},\ }\href {\doibase
  10.1103/RevModPhys.81.109} {\bibfield  {journal} {\bibinfo  {journal} {Rev.
  Mod. Phys.}\ }\textbf {\bibinfo {volume} {81}},\ \bibinfo {pages} {109}
  (\bibinfo {year} {2009})}\BibitemShut {NoStop}%
\bibitem [{\citenamefont {Beenakker}(2008)}]{RevModPhys.80.1337}%
  \BibitemOpen
  \bibfield  {author} {\bibinfo {author} {\bibfnamefont {C.~W.~J.}\
  \bibnamefont {Beenakker}},\ }\href {\doibase 10.1103/RevModPhys.80.1337}
  {\bibfield  {journal} {\bibinfo  {journal} {Rev. Mod. Phys.}\ }\textbf
  {\bibinfo {volume} {80}},\ \bibinfo {pages} {1337} (\bibinfo {year}
  {2008})}\BibitemShut {NoStop}%
\bibitem [{\citenamefont {Peres}(2010)}]{RevModPhys.82.2673}%
  \BibitemOpen
  \bibfield  {author} {\bibinfo {author} {\bibfnamefont {N.~M.~R.}\
  \bibnamefont {Peres}},\ }\href {\doibase 10.1103/RevModPhys.82.2673}
  {\bibfield  {journal} {\bibinfo  {journal} {Rev. Mod. Phys.}\ }\textbf
  {\bibinfo {volume} {82}},\ \bibinfo {pages} {2673} (\bibinfo {year}
  {2010})}\BibitemShut {NoStop}%
\bibitem [{\citenamefont {Zhang}\ and\ \citenamefont
  {Liu}(2008)}]{PhysRevLett.101.264303}%
  \BibitemOpen
  \bibfield  {author} {\bibinfo {author} {\bibfnamefont {X.}~\bibnamefont
  {Zhang}}\ and\ \bibinfo {author} {\bibfnamefont {Z.}~\bibnamefont {Liu}},\
  }\href {\doibase 10.1103/PhysRevLett.101.264303} {\bibfield  {journal}
  {\bibinfo  {journal} {Phys. Rev. Lett.}\ }\textbf {\bibinfo {volume} {101}},\
  \bibinfo {pages} {264303} (\bibinfo {year} {2008})}\BibitemShut {NoStop}%
\bibitem [{\citenamefont {Bittner}\ \emph {et~al.}(2010)\citenamefont
  {Bittner}, \citenamefont {Dietz}, \citenamefont {Miski-Oglu}, \citenamefont
  {Oria~Iriarte}, \citenamefont {Richter},\ and\ \citenamefont
  {Sch\"afer}}]{Bittner2010}%
  \BibitemOpen
  \bibfield  {author} {\bibinfo {author} {\bibfnamefont {S.}~\bibnamefont
  {Bittner}}, \bibinfo {author} {\bibfnamefont {B.}~\bibnamefont {Dietz}},
  \bibinfo {author} {\bibfnamefont {M.}~\bibnamefont {Miski-Oglu}}, \bibinfo
  {author} {\bibfnamefont {P.}~\bibnamefont {Oria~Iriarte}}, \bibinfo {author}
  {\bibfnamefont {A.}~\bibnamefont {Richter}}, \ and\ \bibinfo {author}
  {\bibfnamefont {F.}~\bibnamefont {Sch\"afer}},\ }\href {\doibase
  10.1103/PhysRevB.82.014301} {\bibfield  {journal} {\bibinfo  {journal} {Phys.
  Rev. B}\ }\textbf {\bibinfo {volume} {82}},\ \bibinfo {pages} {014301}
  (\bibinfo {year} {2010})}\BibitemShut {NoStop}%
\bibitem [{\citenamefont {Bittner}\ \emph {et~al.}(2012)\citenamefont
  {Bittner}, \citenamefont {Dietz}, \citenamefont {Miski-Oglu},\ and\
  \citenamefont {Richter}}]{PhysRevB.85.064301}%
  \BibitemOpen
  \bibfield  {author} {\bibinfo {author} {\bibfnamefont {S.}~\bibnamefont
  {Bittner}}, \bibinfo {author} {\bibfnamefont {B.}~\bibnamefont {Dietz}},
  \bibinfo {author} {\bibfnamefont {M.}~\bibnamefont {Miski-Oglu}}, \ and\
  \bibinfo {author} {\bibfnamefont {A.}~\bibnamefont {Richter}},\ }\href
  {\doibase 10.1103/PhysRevB.85.064301} {\bibfield  {journal} {\bibinfo
  {journal} {Phys. Rev. B}\ }\textbf {\bibinfo {volume} {85}},\ \bibinfo
  {pages} {064301} (\bibinfo {year} {2012})}\BibitemShut {NoStop}%
\bibitem [{\citenamefont {Mei}\ \emph {et~al.}(2012)\citenamefont {Mei},
  \citenamefont {Wu}, \citenamefont {Chan},\ and\ \citenamefont
  {Zhang}}]{PhysRevB.86.035141}%
  \BibitemOpen
  \bibfield  {author} {\bibinfo {author} {\bibfnamefont {J.}~\bibnamefont
  {Mei}}, \bibinfo {author} {\bibfnamefont {Y.}~\bibnamefont {Wu}}, \bibinfo
  {author} {\bibfnamefont {C.~T.}\ \bibnamefont {Chan}}, \ and\ \bibinfo
  {author} {\bibfnamefont {Z.-Q.}\ \bibnamefont {Zhang}},\ }\href {\doibase
  10.1103/PhysRevB.86.035141} {\bibfield  {journal} {\bibinfo  {journal} {Phys.
  Rev. B}\ }\textbf {\bibinfo {volume} {86}},\ \bibinfo {pages} {035141}
  (\bibinfo {year} {2012})}\BibitemShut {NoStop}%
\bibitem [{\citenamefont {Ochiai}\ and\ \citenamefont
  {Onoda}(2009)}]{PhysRevB.80.155103}%
  \BibitemOpen
  \bibfield  {author} {\bibinfo {author} {\bibfnamefont {T.}~\bibnamefont
  {Ochiai}}\ and\ \bibinfo {author} {\bibfnamefont {M.}~\bibnamefont {Onoda}},\
  }\href {\doibase 10.1103/PhysRevB.80.155103} {\bibfield  {journal} {\bibinfo
  {journal} {Phys. Rev. B}\ }\textbf {\bibinfo {volume} {80}},\ \bibinfo
  {pages} {155103} (\bibinfo {year} {2009})}\BibitemShut {NoStop}%
\bibitem [{\citenamefont {Zhong}\ and\ \citenamefont
  {Zhang}(2011)}]{zhong2011acoustic}%
  \BibitemOpen
  \bibfield  {author} {\bibinfo {author} {\bibfnamefont {W.}~\bibnamefont
  {Zhong}}\ and\ \bibinfo {author} {\bibfnamefont {X.}~\bibnamefont {Zhang}},\
  }\href@noop {} {\bibfield  {journal} {\bibinfo  {journal} {Physics Letters
  A}\ }\textbf {\bibinfo {volume} {375}},\ \bibinfo {pages} {3533} (\bibinfo
  {year} {2011})}\BibitemShut {NoStop}%
\bibitem [{\citenamefont {Torrent}\ and\ \citenamefont
  {S\'anchez-Dehesa}(2012)}]{PhysRevLett.108.174301}%
  \BibitemOpen
  \bibfield  {author} {\bibinfo {author} {\bibfnamefont {D.}~\bibnamefont
  {Torrent}}\ and\ \bibinfo {author} {\bibfnamefont {J.}~\bibnamefont
  {S\'anchez-Dehesa}},\ }\href {\doibase 10.1103/PhysRevLett.108.174301}
  {\bibfield  {journal} {\bibinfo  {journal} {Phys. Rev. Lett.}\ }\textbf
  {\bibinfo {volume} {108}},\ \bibinfo {pages} {174301} (\bibinfo {year}
  {2012})}\BibitemShut {NoStop}%
\bibitem [{\citenamefont {Torrent}\ \emph {et~al.}(2013)\citenamefont
  {Torrent}, \citenamefont {Mayou},\ and\ \citenamefont
  {S\'anchez-Dehesa}}]{PhysRevB.87.115143}%
  \BibitemOpen
  \bibfield  {author} {\bibinfo {author} {\bibfnamefont {D.}~\bibnamefont
  {Torrent}}, \bibinfo {author} {\bibfnamefont {D.}~\bibnamefont {Mayou}}, \
  and\ \bibinfo {author} {\bibfnamefont {J.}~\bibnamefont {S\'anchez-Dehesa}},\
  }\href {\doibase 10.1103/PhysRevB.87.115143} {\bibfield  {journal} {\bibinfo
  {journal} {Phys. Rev. B}\ }\textbf {\bibinfo {volume} {87}},\ \bibinfo
  {pages} {115143} (\bibinfo {year} {2013})}\BibitemShut {NoStop}%
\bibitem [{\citenamefont {Wallace}(1947)}]{PhysRev.71.622}%
  \BibitemOpen
  \bibfield  {author} {\bibinfo {author} {\bibfnamefont {P.~R.}\ \bibnamefont
  {Wallace}},\ }\href {\doibase 10.1103/PhysRev.71.622} {\bibfield  {journal}
  {\bibinfo  {journal} {Phys. Rev.}\ }\textbf {\bibinfo {volume} {71}},\
  \bibinfo {pages} {622} (\bibinfo {year} {1947})}\BibitemShut {NoStop}%
\bibitem [{\citenamefont {Reich}\ \emph {et~al.}(2002)\citenamefont {Reich},
  \citenamefont {Maultzsch}, \citenamefont {Thomsen},\ and\ \citenamefont
  {Ordej\'on}}]{PhysRevB.66.035412}%
  \BibitemOpen
  \bibfield  {author} {\bibinfo {author} {\bibfnamefont {S.}~\bibnamefont
  {Reich}}, \bibinfo {author} {\bibfnamefont {J.}~\bibnamefont {Maultzsch}},
  \bibinfo {author} {\bibfnamefont {C.}~\bibnamefont {Thomsen}}, \ and\
  \bibinfo {author} {\bibfnamefont {P.}~\bibnamefont {Ordej\'on}},\ }\href
  {\doibase 10.1103/PhysRevB.66.035412} {\bibfield  {journal} {\bibinfo
  {journal} {Phys. Rev. B}\ }\textbf {\bibinfo {volume} {66}},\ \bibinfo
  {pages} {035412} (\bibinfo {year} {2002})}\BibitemShut {NoStop}%
\bibitem [{\citenamefont {Kuhl}\ \emph {et~al.}(2010)\citenamefont {Kuhl},
  \citenamefont {Barkhofen}, \citenamefont {Tudorovskiy}, \citenamefont
  {St\"ockmann}, \citenamefont {Hossain}, \citenamefont {de~Forges~de Parny},\
  and\ \citenamefont {Mortessagne}}]{PhysRevB.82.094308}%
  \BibitemOpen
  \bibfield  {author} {\bibinfo {author} {\bibfnamefont {U.}~\bibnamefont
  {Kuhl}}, \bibinfo {author} {\bibfnamefont {S.}~\bibnamefont {Barkhofen}},
  \bibinfo {author} {\bibfnamefont {T.}~\bibnamefont {Tudorovskiy}}, \bibinfo
  {author} {\bibfnamefont {H.-J.}\ \bibnamefont {St\"ockmann}}, \bibinfo
  {author} {\bibfnamefont {T.}~\bibnamefont {Hossain}}, \bibinfo {author}
  {\bibfnamefont {L.}~\bibnamefont {de~Forges~de Parny}}, \ and\ \bibinfo
  {author} {\bibfnamefont {F.}~\bibnamefont {Mortessagne}},\ }\href {\doibase
  10.1103/PhysRevB.82.094308} {\bibfield  {journal} {\bibinfo  {journal} {Phys.
  Rev. B}\ }\textbf {\bibinfo {volume} {82}},\ \bibinfo {pages} {094308}
  (\bibinfo {year} {2010})}\BibitemShut {NoStop}%
\bibitem [{\citenamefont {Bellec}\ \emph {et~al.}(2013)\citenamefont {Bellec},
  \citenamefont {Kuhl}, \citenamefont {Montambaux},\ and\ \citenamefont
  {Mortessagne}}]{PhysRevB.88.115437}%
  \BibitemOpen
  \bibfield  {author} {\bibinfo {author} {\bibfnamefont {M.}~\bibnamefont
  {Bellec}}, \bibinfo {author} {\bibfnamefont {U.}~\bibnamefont {Kuhl}},
  \bibinfo {author} {\bibfnamefont {G.}~\bibnamefont {Montambaux}}, \ and\
  \bibinfo {author} {\bibfnamefont {F.}~\bibnamefont {Mortessagne}},\ }\href
  {\doibase 10.1103/PhysRevB.88.115437} {\bibfield  {journal} {\bibinfo
  {journal} {Phys. Rev. B}\ }\textbf {\bibinfo {volume} {88}},\ \bibinfo
  {pages} {115437} (\bibinfo {year} {2013})}\BibitemShut {NoStop}%
\bibitem [{\citenamefont {Fang}\ \emph {et~al.}(2011)\citenamefont {Fang},
  \citenamefont {Yu},\ and\ \citenamefont {Fan}}]{PhysRevB.84.075477}%
  \BibitemOpen
  \bibfield  {author} {\bibinfo {author} {\bibfnamefont {K.}~\bibnamefont
  {Fang}}, \bibinfo {author} {\bibfnamefont {Z.}~\bibnamefont {Yu}}, \ and\
  \bibinfo {author} {\bibfnamefont {S.}~\bibnamefont {Fan}},\ }\href {\doibase
  10.1103/PhysRevB.84.075477} {\bibfield  {journal} {\bibinfo  {journal} {Phys.
  Rev. B}\ }\textbf {\bibinfo {volume} {84}},\ \bibinfo {pages} {075477}
  (\bibinfo {year} {2011})}\BibitemShut {NoStop}%
\bibitem [{\citenamefont {Han}\ \emph {et~al.}(2009)\citenamefont {Han},
  \citenamefont {Lai}, \citenamefont {Zi}, \citenamefont {Zhang},\ and\
  \citenamefont {Chan}}]{PhysRevLett.102.123904}%
  \BibitemOpen
  \bibfield  {author} {\bibinfo {author} {\bibfnamefont {D.}~\bibnamefont
  {Han}}, \bibinfo {author} {\bibfnamefont {Y.}~\bibnamefont {Lai}}, \bibinfo
  {author} {\bibfnamefont {J.}~\bibnamefont {Zi}}, \bibinfo {author}
  {\bibfnamefont {Z.-Q.}\ \bibnamefont {Zhang}}, \ and\ \bibinfo {author}
  {\bibfnamefont {C.~T.}\ \bibnamefont {Chan}},\ }\href {\doibase
  10.1103/PhysRevLett.102.123904} {\bibfield  {journal} {\bibinfo  {journal}
  {Phys. Rev. Lett.}\ }\textbf {\bibinfo {volume} {102}},\ \bibinfo {pages}
  {123904} (\bibinfo {year} {2009})}\BibitemShut {NoStop}%
\bibitem [{\citenamefont {Engheta}(2006)}]{Engheta2006}%
  \BibitemOpen
  \bibfield  {author} {\bibinfo {author} {\bibfnamefont {R.}~\bibnamefont
  {Engheta}, \bibfnamefont {N.~\&~Ziolkowski}},\ }\href@noop {} {\emph
  {\bibinfo {title} {Metamaterials: Physics and Engineering Explorations
  (Wiley, IEEE Press, 2006)}}},\ edited by\ \bibinfo {editor} {\bibfnamefont
  {I.}~\bibnamefont {Press}}\ (\bibinfo  {publisher} {Wiley},\ \bibinfo {year}
  {2006})\BibitemShut {NoStop}%
\bibitem [{\citenamefont {Deymier}(2013)}]{deymier2013acoustic}%
  \BibitemOpen
  \bibfield  {author} {\bibinfo {author} {\bibfnamefont {P.~A.}\ \bibnamefont
  {Deymier}},\ }\href@noop {} {\emph {\bibinfo {title} {Acoustic metamaterials
  and phononic crystals}}},\ Vol.\ \bibinfo {volume} {173}\ (\bibinfo
  {publisher} {Springer Science \& Business Media},\ \bibinfo {year}
  {2013})\BibitemShut {NoStop}%
\bibitem [{\citenamefont {Cummer}\ \emph {et~al.}(2016)\citenamefont {Cummer},
  \citenamefont {Christensen},\ and\ \citenamefont {Al{\`u}}}]{Cummer2016}%
  \BibitemOpen
  \bibfield  {author} {\bibinfo {author} {\bibfnamefont {S.~A.}\ \bibnamefont
  {Cummer}}, \bibinfo {author} {\bibfnamefont {J.}~\bibnamefont {Christensen}},
  \ and\ \bibinfo {author} {\bibfnamefont {A.}~\bibnamefont {Al{\`u}}},\ }\href
  {http://dx.doi.org/10.1038/natrevmats.2016.1} {\bibfield  {journal} {\bibinfo
   {journal} {Nature Reviews Materials}\ }\textbf {\bibinfo {volume} {1}},\
  \bibinfo {pages} {16001 EP } (\bibinfo {year} {2016})}\BibitemShut {NoStop}%
\bibitem [{\citenamefont {Ma}\ and\ \citenamefont {Sheng}(2016)}]{Mae1501595}%
  \BibitemOpen
  \bibfield  {author} {\bibinfo {author} {\bibfnamefont {G.}~\bibnamefont
  {Ma}}\ and\ \bibinfo {author} {\bibfnamefont {P.}~\bibnamefont {Sheng}},\
  }\href {\doibase 10.1126/sciadv.1501595} {\bibfield  {journal} {\bibinfo
  {journal} {Science Advances}\ }\textbf {\bibinfo {volume} {2}} (\bibinfo
  {year} {2016}),\ 10.1126/sciadv.1501595}\BibitemShut {NoStop}%
\bibitem [{\citenamefont {Li}\ and\ \citenamefont
  {Chan}(2004)}]{PhysRevE.70.055602}%
  \BibitemOpen
  \bibfield  {author} {\bibinfo {author} {\bibfnamefont {J.}~\bibnamefont
  {Li}}\ and\ \bibinfo {author} {\bibfnamefont {C.~T.}\ \bibnamefont {Chan}},\
  }\href {\doibase 10.1103/PhysRevE.70.055602} {\bibfield  {journal} {\bibinfo
  {journal} {Phys. Rev. E}\ }\textbf {\bibinfo {volume} {70}},\ \bibinfo
  {pages} {055602} (\bibinfo {year} {2004})}\BibitemShut {NoStop}%
\bibitem [{\citenamefont {Xie}\ \emph {et~al.}(2013)\citenamefont {Xie},
  \citenamefont {Popa}, \citenamefont {Zigoneanu},\ and\ \citenamefont
  {Cummer}}]{PhysRevLett.110.175501}%
  \BibitemOpen
  \bibfield  {author} {\bibinfo {author} {\bibfnamefont {Y.}~\bibnamefont
  {Xie}}, \bibinfo {author} {\bibfnamefont {B.-I.}\ \bibnamefont {Popa}},
  \bibinfo {author} {\bibfnamefont {L.}~\bibnamefont {Zigoneanu}}, \ and\
  \bibinfo {author} {\bibfnamefont {S.~A.}\ \bibnamefont {Cummer}},\ }\href
  {\doibase 10.1103/PhysRevLett.110.175501} {\bibfield  {journal} {\bibinfo
  {journal} {Phys. Rev. Lett.}\ }\textbf {\bibinfo {volume} {110}},\ \bibinfo
  {pages} {175501} (\bibinfo {year} {2013})}\BibitemShut {NoStop}%
\bibitem [{\citenamefont {Zigoneanu}\ \emph {et~al.}(2014)\citenamefont
  {Zigoneanu}, \citenamefont {Popa},\ and\ \citenamefont
  {Cummer}}]{Zigoneanu2014}%
  \BibitemOpen
  \bibfield  {author} {\bibinfo {author} {\bibfnamefont {L.}~\bibnamefont
  {Zigoneanu}}, \bibinfo {author} {\bibfnamefont {B.-I.}\ \bibnamefont {Popa}},
  \ and\ \bibinfo {author} {\bibfnamefont {S.~A.}\ \bibnamefont {Cummer}},\
  }\href {http://dx.doi.org/10.1038/nmat3901} {\bibfield  {journal} {\bibinfo
  {journal} {Nat Mater}\ }\textbf {\bibinfo {volume} {13}},\ \bibinfo {pages}
  {352} (\bibinfo {year} {2014})}\BibitemShut {NoStop}%
\bibitem [{\citenamefont {Garc\'{\i}a-Chocano}\ \emph
  {et~al.}(2014)\citenamefont {Garc\'{\i}a-Chocano}, \citenamefont
  {Christensen},\ and\ \citenamefont
  {S\'anchez-Dehesa}}]{PhysRevLett.112.144301}%
  \BibitemOpen
  \bibfield  {author} {\bibinfo {author} {\bibfnamefont {V.~M.}\ \bibnamefont
  {Garc\'{\i}a-Chocano}}, \bibinfo {author} {\bibfnamefont {J.}~\bibnamefont
  {Christensen}}, \ and\ \bibinfo {author} {\bibfnamefont {J.}~\bibnamefont
  {S\'anchez-Dehesa}},\ }\href {\doibase 10.1103/PhysRevLett.112.144301}
  {\bibfield  {journal} {\bibinfo  {journal} {Phys. Rev. Lett.}\ }\textbf
  {\bibinfo {volume} {112}},\ \bibinfo {pages} {144301} (\bibinfo {year}
  {2014})}\BibitemShut {NoStop}%
\bibitem [{\citenamefont {Christensen}\ and\ \citenamefont
  {de~Abajo}(2012)}]{PhysRevLett.108.124301}%
  \BibitemOpen
  \bibfield  {author} {\bibinfo {author} {\bibfnamefont {J.}~\bibnamefont
  {Christensen}}\ and\ \bibinfo {author} {\bibfnamefont {F.~J.~G.}\
  \bibnamefont {de~Abajo}},\ }\href {\doibase 10.1103/PhysRevLett.108.124301}
  {\bibfield  {journal} {\bibinfo  {journal} {Phys. Rev. Lett.}\ }\textbf
  {\bibinfo {volume} {108}},\ \bibinfo {pages} {124301} (\bibinfo {year}
  {2012})}\BibitemShut {NoStop}%
\bibitem [{\citenamefont {Fleury}\ and\ \citenamefont
  {Al{\`u}}(2013)}]{fleury2013extraordinary}%
  \BibitemOpen
  \bibfield  {author} {\bibinfo {author} {\bibfnamefont {R.}~\bibnamefont
  {Fleury}}\ and\ \bibinfo {author} {\bibfnamefont {A.}~\bibnamefont
  {Al{\`u}}},\ }\href@noop {} {\bibfield  {journal} {\bibinfo  {journal}
  {Physical review letters}\ }\textbf {\bibinfo {volume} {111}},\ \bibinfo
  {pages} {055501} (\bibinfo {year} {2013})}\BibitemShut {NoStop}%
\bibitem [{\citenamefont {Zhu}\ \emph {et~al.}(2011)\citenamefont {Zhu},
  \citenamefont {Christensen}, \citenamefont {Jung}, \citenamefont
  {Martin-Moreno}, \citenamefont {Yin}, \citenamefont {Fok}, \citenamefont
  {Zhang},\ and\ \citenamefont {Garcia-Vidal}}]{Zhu2011}%
  \BibitemOpen
  \bibfield  {author} {\bibinfo {author} {\bibfnamefont {J.}~\bibnamefont
  {Zhu}}, \bibinfo {author} {\bibfnamefont {J.}~\bibnamefont {Christensen}},
  \bibinfo {author} {\bibfnamefont {J.}~\bibnamefont {Jung}}, \bibinfo {author}
  {\bibfnamefont {L.}~\bibnamefont {Martin-Moreno}}, \bibinfo {author}
  {\bibfnamefont {X.}~\bibnamefont {Yin}}, \bibinfo {author} {\bibfnamefont
  {L.}~\bibnamefont {Fok}}, \bibinfo {author} {\bibfnamefont {X.}~\bibnamefont
  {Zhang}}, \ and\ \bibinfo {author} {\bibfnamefont {F.~J.}\ \bibnamefont
  {Garcia-Vidal}},\ }\href {\doibase 10.1038/nphys1804} {\bibfield  {journal}
  {\bibinfo  {journal} {Nat Phys}\ }\textbf {\bibinfo {volume} {7}},\ \bibinfo
  {pages} {52} (\bibinfo {year} {2011})}\BibitemShut {NoStop}%
\bibitem [{\citenamefont {Liu}\ \emph {et~al.}(2000)\citenamefont {Liu},
  \citenamefont {Zhang}, \citenamefont {Mao}, \citenamefont {Zhu},
  \citenamefont {Yang}, \citenamefont {Chan},\ and\ \citenamefont
  {Sheng}}]{Liu1734}%
  \BibitemOpen
  \bibfield  {author} {\bibinfo {author} {\bibfnamefont {Z.}~\bibnamefont
  {Liu}}, \bibinfo {author} {\bibfnamefont {X.}~\bibnamefont {Zhang}}, \bibinfo
  {author} {\bibfnamefont {Y.}~\bibnamefont {Mao}}, \bibinfo {author}
  {\bibfnamefont {Y.~Y.}\ \bibnamefont {Zhu}}, \bibinfo {author} {\bibfnamefont
  {Z.}~\bibnamefont {Yang}}, \bibinfo {author} {\bibfnamefont {C.~T.}\
  \bibnamefont {Chan}}, \ and\ \bibinfo {author} {\bibfnamefont
  {P.}~\bibnamefont {Sheng}},\ }\href {\doibase 10.1126/science.289.5485.1734}
  {\bibfield  {journal} {\bibinfo  {journal} {Science}\ }\textbf {\bibinfo
  {volume} {289}},\ \bibinfo {pages} {1734} (\bibinfo {year}
  {2000})}\BibitemShut {NoStop}%
\bibitem [{\citenamefont {Achilleos}\ \emph {et~al.}(2015)\citenamefont
  {Achilleos}, \citenamefont {Richoux}, \citenamefont {Theocharis},\ and\
  \citenamefont {Frantzeskakis}}]{PhysRevE.91.023204}%
  \BibitemOpen
  \bibfield  {author} {\bibinfo {author} {\bibfnamefont {V.}~\bibnamefont
  {Achilleos}}, \bibinfo {author} {\bibfnamefont {O.}~\bibnamefont {Richoux}},
  \bibinfo {author} {\bibfnamefont {G.}~\bibnamefont {Theocharis}}, \ and\
  \bibinfo {author} {\bibfnamefont {D.~J.}\ \bibnamefont {Frantzeskakis}},\
  }\href {\doibase 10.1103/PhysRevE.91.023204} {\bibfield  {journal} {\bibinfo
  {journal} {Phys. Rev. E}\ }\textbf {\bibinfo {volume} {91}},\ \bibinfo
  {pages} {023204} (\bibinfo {year} {2015})}\BibitemShut {NoStop}%
\bibitem [{\citenamefont {Romero-Garc{\'i}a}\ \emph {et~al.}(2016)\citenamefont
  {Romero-Garc{\'i}a}, \citenamefont {Theocharis}, \citenamefont {Richoux},
  \citenamefont {Merkel}, \citenamefont {Tournat},\ and\ \citenamefont
  {Pagneux}}]{Romero-Garcia2016}%
  \BibitemOpen
  \bibfield  {author} {\bibinfo {author} {\bibfnamefont {V.}~\bibnamefont
  {Romero-Garc{\'i}a}}, \bibinfo {author} {\bibfnamefont {G.}~\bibnamefont
  {Theocharis}}, \bibinfo {author} {\bibfnamefont {O.}~\bibnamefont {Richoux}},
  \bibinfo {author} {\bibfnamefont {A.}~\bibnamefont {Merkel}}, \bibinfo
  {author} {\bibfnamefont {V.}~\bibnamefont {Tournat}}, \ and\ \bibinfo
  {author} {\bibfnamefont {V.}~\bibnamefont {Pagneux}},\ }\href
  {http://dx.doi.org/10.1038/srep19519} {\bibfield  {journal} {\bibinfo
  {journal} {Scientific Reports}\ }\textbf {\bibinfo {volume} {6}},\ \bibinfo
  {pages} {19519 EP } (\bibinfo {year} {2016})}\BibitemShut {NoStop}%
\bibitem [{\citenamefont {Lemoult}\ \emph {et~al.}(2011)\citenamefont
  {Lemoult}, \citenamefont {Fink},\ and\ \citenamefont
  {Lerosey}}]{PhysRevLett.107.064301}%
  \BibitemOpen
  \bibfield  {author} {\bibinfo {author} {\bibfnamefont {F.}~\bibnamefont
  {Lemoult}}, \bibinfo {author} {\bibfnamefont {M.}~\bibnamefont {Fink}}, \
  and\ \bibinfo {author} {\bibfnamefont {G.}~\bibnamefont {Lerosey}},\ }\href
  {\doibase 10.1103/PhysRevLett.107.064301} {\bibfield  {journal} {\bibinfo
  {journal} {Phys. Rev. Lett.}\ }\textbf {\bibinfo {volume} {107}},\ \bibinfo
  {pages} {064301} (\bibinfo {year} {2011})}\BibitemShut {NoStop}%
\bibitem [{\citenamefont {Lemoult}\ \emph {et~al.}(2013)\citenamefont
  {Lemoult}, \citenamefont {Kaina}, \citenamefont {Fink},\ and\ \citenamefont
  {Lerosey}}]{Lemoult2013}%
  \BibitemOpen
  \bibfield  {author} {\bibinfo {author} {\bibfnamefont {F.}~\bibnamefont
  {Lemoult}}, \bibinfo {author} {\bibfnamefont {N.}~\bibnamefont {Kaina}},
  \bibinfo {author} {\bibfnamefont {M.}~\bibnamefont {Fink}}, \ and\ \bibinfo
  {author} {\bibfnamefont {G.}~\bibnamefont {Lerosey}},\ }\href {\doibase
  10.1038/nphys2480} {\bibfield  {journal} {\bibinfo  {journal} {Nat Phys}\
  }\textbf {\bibinfo {volume} {9}},\ \bibinfo {pages} {55} (\bibinfo {year}
  {2013})}\BibitemShut {NoStop}%
\bibitem [{\citenamefont {Kaina}\ \emph {et~al.}(2015)\citenamefont {Kaina},
  \citenamefont {Lemoult}, \citenamefont {Fink},\ and\ \citenamefont
  {Lerosey}}]{Kaina2015}%
  \BibitemOpen
  \bibfield  {author} {\bibinfo {author} {\bibfnamefont {N.}~\bibnamefont
  {Kaina}}, \bibinfo {author} {\bibfnamefont {F.}~\bibnamefont {Lemoult}},
  \bibinfo {author} {\bibfnamefont {M.}~\bibnamefont {Fink}}, \ and\ \bibinfo
  {author} {\bibfnamefont {G.}~\bibnamefont {Lerosey}},\ }\href
  {http://dx.doi.org/10.1038/nature14678} {\bibfield  {journal} {\bibinfo
  {journal} {Nature}\ }\textbf {\bibinfo {volume} {525}},\ \bibinfo {pages}
  {77} (\bibinfo {year} {2015})}\BibitemShut {NoStop}%
\bibitem [{\citenamefont {Lemoult}\ \emph {et~al.}(2016)\citenamefont
  {Lemoult}, \citenamefont {Kaina}, \citenamefont {Fink},\ and\ \citenamefont
  {Lerosey}}]{Lemoult_2016}%
  \BibitemOpen
  \bibfield  {author} {\bibinfo {author} {\bibfnamefont {F.}~\bibnamefont
  {Lemoult}}, \bibinfo {author} {\bibfnamefont {N.}~\bibnamefont {Kaina}},
  \bibinfo {author} {\bibfnamefont {M.}~\bibnamefont {Fink}}, \ and\ \bibinfo
  {author} {\bibfnamefont {G.}~\bibnamefont {Lerosey}},\ }\href {\doibase
  10.3390/cryst6070082} {\bibfield  {journal} {\bibinfo  {journal} {Crystals}\
  }\textbf {\bibinfo {volume} {6}},\ \bibinfo {pages} {82} (\bibinfo {year}
  {2016})}\BibitemShut {NoStop}%
\bibitem [{\citenamefont {Sigalas}\ and\ \citenamefont
  {Economou}(1993)}]{sigalas1993band}%
  \BibitemOpen
  \bibfield  {author} {\bibinfo {author} {\bibfnamefont {M.}~\bibnamefont
  {Sigalas}}\ and\ \bibinfo {author} {\bibfnamefont {E.~N.}\ \bibnamefont
  {Economou}},\ }\href@noop {} {\bibfield  {journal} {\bibinfo  {journal}
  {Solid State Communications}\ }\textbf {\bibinfo {volume} {86}},\ \bibinfo
  {pages} {141} (\bibinfo {year} {1993})}\BibitemShut {NoStop}%
\end{thebibliography}%

\end{document}